\documentclass[namedreferences]{kluwer}

\usepackage{epsfig}

\begin{document}

\begin{article}

\begin{opening}

\title{A method of self-consistent three-dimensional solution of the
magnetohydrostatic equations using magnetogram data}

\author{G. V. \surname{Rudenko}\email{rud@iszf.irk.ru}}
\institute{Institute of Solar-Terrestrial Physics, Irkutsk, Russia}

\runningtitle{A method of self-consistent three-dimensional solution}

\runningauthor{G.V. RUDENKO}

\begin{ao}
  Institute of Solar-Terrestrial Physics,
  P.O. Box 4026, Irkutsk, 664033, Russia \\
  e-mail: rud@iszf.irk.ru \\
  Fax: +7 (3952) 46 25 57
\end{ao}


\begin{abstract}
A technique is proposed for constructing three-dimensional solutions
comlplying with the self-consistent magnetohydrostatic (MHS) equations and
with observations along the line of sight of the magnetic field at the
photosphere. The technique is a generalization of a paper \cite{Rudenko} to
a potential-field approximation. The solution of the problem under
consideration leans a representation of the magnetic field in terms of a
scalar function, with its subsequent harmonic expansion in terms of a
spherical functional basis that satisfies specified boundary conditions. A
numerical realization of the proposed method is expected to permit a
real-time modelling of three-dimensional magnetic field, temperature,
pressure and density distributions.
\end{abstract}

\end{opening}

\section{Introduction}

In a paper \cite{Rudenko} it was suggested a method to model a
three-dimensional distribution of the above-photospheric magnetic field,
base on solving the Laplace boundary-value problem that corresponds to
actual one-component magnetic measurements at the photospheric level.
Boundary conditions on the lower sphere were specified in \cite{Rudenko} by
a distribution of the projection of the magnetic field onto a certain fixed
direction corresponding to the line of sight at the time of measurement. In
this paper, the same boundary conditions are used in solving a
boundary-value problem for a certain class of self-consistent
magnetohydrostatic (MHS) equations admitting of analytic solutions.
Developing self-consistent physical models of chromospheric and coronal
plasmas is one of the major tasks of theoretical astrophysics. Using the MHS
approximation is justified when it is intended to study the current state of
the medium or relatively slowly occurring macroscopic plasma processes.
Also, because of the non-linearity of the MHS equations, the problem of
seeking all possible solutions is extremely complicated. Several approaches
to constructing self-consistent MHS models are known to date, which permit
analytic solutions to be determined. These approaches are based on some not
fundamentally differing preliminary assumptions about the way in which the
solutions behave. An outline of the approaches and research results on this
subject may be found in (\cite{Low1},\cite{Low2} , \cite{Low3}, \cite{Low4},
\cite{Low5}, \cite{Bogdan}, and \cite{N}). The class of the models studied
thus far is relatively narrow to pretend to a generality of the description
of the real states of magnetic plasma. Nevertheless, these models, when
used, would be expected to give at least qualitative insight into the
possible correlations between the fundamental physical plasma
characteristics. There is no question that the MHS modelling is much higher
in information content than the simplest modelling in the potential-field
approximation of the magnetic field in extensive use to date.

An important factor as to the extrapolation of the magnetic field and other
physical characteristics in terms of a particular physical model is the
ability to derive physical solutions that satisfy boundary conditions
corresponding to actual measurements. Only then can we expect that the
result of an extrapolation will represent the facts, and only then can we
select -- on the basis of many comparisons -- the most adequate physical
model for use in the study. Three methods of specifying boundary conditions
are customarily used to extrapolate magnetic fields in the potential-field
approximation: from the radial component $B_r$ (Neumann's classical
boundary-value problem); from the component $B_l$ (projection along the
direction corresponding to the orientation of the line of sight at the time
when each point crosses the surface of the central meridian:, and from the
component $B_d$ (projection along the line of sight corresponding to the
time of measurement \cite{Rudenko}). Only the second and third methods of
specifying boundary conditions correspond to actual magnetic field
observations. More specifically, the second method gives only an averaged
picture of the magnetic configuration (to accumulate boundary conditions
requires a full series of daily observations over the course of a complete
rotation). The third method appears to be the most correct as only its
application can provide the best fit to a current state of the magnetic
field. As in the case of the potential-field approximation, of important
significance is the statement and solution of a boundary-value problem in
the MHS approximation with boundary conditions of the third type (from the
component $B_d$). In this paper such a setting of the problem is realized
for MHS models as suggested in \cite{N} where analytic solutions were
obtained for MHS models satisfying boundary conditions of the 1st type (from
the component $B_r$). In this paper we consider the same MHS approximations,
yet for different boundary conditions. We restrict the discussion only to
the analytic solution of the problem formulated. It can be used in
constructing a numerical algorithm for calculating a current state of solar
plasma, and this issue will be the subject of further investigation. The
problems associated with scarcity of data from the averted side of the Sun
will not be discussed here as they were covered in considerable detail in a
previous paper \cite{Rudenko} for the potential-field case.

\section{Setting up the problem, and the method of solution.}

Here we take advantage of the physical models whose substantiation and
physical aspects are reported in \cite{N}. We restrict ourselves only to a
brief formulation of the problem in the form of equations. From the logical
standpoint, the problem is conveniently broken up into three parts: one is
the system of basic equations; the other involves imposing additional
requirements on the function of current density (strictly speaking, a
concrete definition of the physical model); and the third includes
specifying the boundary conditions. The system of MHS equations is

\begin{equation}
{\bf j}\times {\bf B}-{\bf \nabla }p-\rho {\bf \nabla }\psi ={\bf 0,}
\label{b1}
\end{equation}
\begin{equation}
{\bf \nabla }\times {\bf B}=\mu _0{\bf j,}  \label{b2}
\end{equation}
\begin{equation}
{\bf \nabla }\cdot {\bf B}=0.  \label{b3}
\end{equation}
Here ${\bf B}$, ${\bf j}${\bf ,} $p$, $\rho $and $\psi $ are magnetic
induction, current density, pressure, density, and gravitational potential,
respectively.

As shown in \cite{Low2} and \cite{N}, the current density that allows
pressure and density from (\ref{b1}) to be expressed explicitly in terms of
the magnetic field has the form:

\begin{equation}
{\bf j}=\alpha {\bf B}+{\bf \nabla }F\left( {\bf \nabla }\psi \cdot {\bf B}%
,\psi \right) \times {\bf \nabla }\psi ,  \label{b4}
\end{equation}
where $F$ is an arbitrary function of its arguments. As is \cite{N}, we
consider a simplified model: $\alpha =const$, and

\begin{equation}
F=k(\psi )\frac{GM}{r^3}{\bf r}\cdot {\bf B=}\xi {\bf (}r{\bf )}\frac{r^3}{GM%
}{\bf r\cdot B,\qquad }\xi {\bf (}r{\bf )=}k(\psi )\left( \frac{GM}{r^3}%
\right) ^2.  \label{b5}
\end{equation}
Here the gravitational potential is specified as

\begin{equation}
\psi =-\frac{GM}r{\bf ,\qquad \nabla }\psi =-\frac{GM}{r^3}{\bf r,}
\label{b6}
\end{equation}
with $M$ being the mass of the central body. For ${\bf \xi (}r{\bf )}$, the
same variants as in (\cite{N}, (40)) can be considered

\begin{equation}
\mu _0\xi {\bf (}r{\bf )}=\left\{
\begin{array}{c}
k/r^2,{\bf \qquad \qquad \qquad \qquad \qquad }case{\bf \ I,} \\
1/r^2-1/(r+a)^2,\qquad \quad \quad case{\bf \ II,} \\
1/r^2+k\exp (-2r/L)-q,\quad case{\bf \ III,}
\end{array}
\right.  \label{b7}
\end{equation}
where $k$, $q$, and $L$ are constants.

To formulate the boundary conditions we introduced in the direction selected
a fixed unit vector ${\bf d}$, corresponding to the direction of the line of
sight, and require the fulfillment of the conditions

\begin{equation}
{\bf d\cdot B}_{|_{r=R}}=B_d(\theta ,\phi ),  \label{b8}
\end{equation}
\begin{equation}
{\bf B}_{|_{r\rightarrow \infty }}\rightarrow 0.  \label{b9}
\end{equation}
Here $B_d(\theta ,\phi )$ is some specified distribution on a spherical
surface of radius $R$.

Formulas (\ref{b2}-\ref{b9}) are sufficient for a full determination of the
magnetic field in the region of space $r>R$. The hydrostatic balance
equation (\ref{b1}) describes pressure and density variations associated
with the magnetic field. At this point we shall not give the formulas
describing the specific relation of these quantities to the magnetic field;
their derivation and particular representation may be found in (\cite{N},
(25)-(31)). In (\cite{N}, see (24)), an analytic expression was obtained for
the magnetic field in terms of the coefficient of spherical harmonic
expansion of the scalar product $({\bf r}\cdot {\bf B)}$. It can be
demonstrated that the magnetic field satisfying (\ref{b2}-\ref{b5}) may be
represented in a more general (than in \cite{N}) form in terms of a single
scalar function. It is easy to verify by mere substitution that the magnetic
field that is expressed as

\begin{equation}
{\bf B=}\left\{ -{\bf r}\triangle +\frac{\overline{\alpha }}i{\bf L}+{\bf %
\nabla }\left( \frac{{\bf r}}r\cdot {\bf \nabla }\right) r\right\} \chi
\label{b10}
\end{equation}
in terms of a certain scalar function $\chi $ satisfying the equation

\begin{equation}
\left( \triangle +\overline{\xi }{\bf L}^2+\overline{\alpha }^2\right) \chi
=0  \label{b11}
\end{equation}
satisfies equations (\ref{b2}-\ref{b5}). In (\ref{b10}, \ref{b11}), $%
\overline{\alpha }=\mu _0\alpha $, $\overline{\xi }=\mu _0\xi $, and ${\bf L}
$ is the angular moment operator (see \cite{N})

\begin{equation}
{\bf L}=\frac 1i{\bf r\times \nabla .}  \label{b12}
\end{equation}
Note that equation (\ref{b11}) for $\chi $ is identical to the equation for $%
({\bf r}\cdot {\bf B)}$ in \cite{N}. Hence the initial problem can be
reformulated as a boundary-value problem for $\chi $, satisfying (\ref{b11})
and the boundary conditions (\ref{b8}), (\ref{b9}).

To solve the boundary-value problem formulated we introduce a spherical
system of coordinates $(r,\theta ,\phi )$ connected with the unit vector $%
{\bf d}$ (as done in \cite{Rudenko}). Let the axis $z$ that correspond to
the selected spherical system of coordinate, be directed along ${\bf d}$,
and let the axis $x$, for definiteness, be taken to lie in the plane
produced by the axis $z$ and the heliographic axis $z^{\prime }$. Finally,
let the function $\chi $ be represented as an expansion into a series in
terms of spherical harmonics of the function

\begin{equation}
\Psi \left( r,\theta ,\phi \right) =R\sum\limits_{l=1}^{\infty \vee
L}\sum\limits_{m=-l}^lc_l^mf_l(r)\widetilde{P}_l^{|m|}\left( \cos \theta
\right) e^{im\phi },  \label{b13}
\end{equation}
where $c_l^m$ are the desired complex expansion coefficients

\begin{equation}
c_l^{-m}=\overline{c}_l^m,  \label{b14}
\end{equation}
$\widetilde{P}_l^m=P_l^m/\sqrt{2\pi w_l^m}$, $P_l^m$ is the Legendre
function (see \cite{Abramowitz and Stegun}), and

\begin{equation}
w_l^m=\int\limits_{-1}^1\left[ P_l^m\left( u\right) \right]
^2du=(l+1/2)^{-1}(l+m)!/(l-m)!.  \label{b15}
\end{equation}
The expression (\ref{b13}) should be regarded as an exact solution if the
summation with respect to the index $l$ is made ad infinitum, or as a finite
approximation if the summation is limited to the value of $L$ (the main
index of expansion). Substitution of (\ref{b13}) into equation (\ref{b11})
gives an equation which must be satisfied by the function $f_l(r)$:

\begin{equation}
\left( \frac{d^2}{dr^2}-l(l+1)\left( \frac 1{r^2}-\overline{\xi }(r)\right) +%
\overline{\alpha }\right) rf_l(r)=0.  \label{b16}
\end{equation}
For the sake of simplicity, we did not introduced one further summation into
(\ref{b13}) - with respect to the indices of two independent solutions of
equation (\ref{b16}). For the three cases of the dependences $\overline{\xi }%
(r)$ in (\ref{b7}), the solitons of equation (\ref{b16}) are expressed in
terms of different variants of Bessel functions (for the explicit expression
see in (\cite{N}, (40)-(44)). Out of them we may always choose linear
combinations satisfying the upper boundary conditions (\ref{b9}). Moreover,
for the case of $\alpha =0$ we can impose on some surface $|r|=R_w>R$ the
boundary conditions of a radial magnetic field. Such conditions are used
very widely in the technique for extrapolating a potential magnetic field in
modelling the conditions on the source surface. As will be apparent from the
ultimate expansion for a full magnetic field vector, such conditions can be
obtained by satisfying the equality

\begin{equation}
\frac d{dr}r\left( a_1f_l^{(1)}(r)+a_2f_l^{(2)}(r)\right) =0,\qquad
f_l(r)=a_1f_l^{(1)}(r)+a_2f_l^{(2)}(r).  \label{b17}
\end{equation}
When $\alpha \neq 0$ such conditions cannot be satisfied in principle;
however, if $\alpha $ is small, the condition (\ref{b17}) might still be
used. In this case the field will be approximately a radial one. In any
case, here $f_l(r)$ is taken to mean the solutions of equation (\ref{b16}),
satisfying some of the upper boundary conditions. Explicit expressions for
them can always be obtained. Upon substituting the expansion (\ref{b13})
into the expression for the magnetic field (\ref{b10}), we arrive at the
expansion of the full magnetic field vector

\begin{equation}
{\bf B(r)}=R\sum\limits_{l=1}^{\infty \vee
L}\sum\limits_{m=-l}^lc_l^m\left\{ \frac{{\bf r}}{r^2}l(l+1)+\frac{\overline{%
\alpha }}i{\bf L}+{\bf \nabla }_{\perp }\frac d{dr}r\right\} f_l\widetilde{P}%
_l^{|m|}(\cos \theta )e^{im\phi }.  \label{b18}
\end{equation}
This expansion coincides exactly in its form with the expansion in (\cite{N}%
, (24)). It can be used to calculate the magnetic field once its
coefficients are determined. Note that by virtue of the identity $({\bf %
r\cdot B)}={\bf L}^2\chi $, the coefficients $c_l^m$ are those of the
initial expansion $({\bf r\cdot B)}$ that is represented by the expression (%
\cite{N}, (17)).

To solve the boundary-value problem formulated, an explicit form of the
expansion of the component $({\bf d\cdot B)}$ is required. Using the
representation of the operators ${\bf L}$ and $\nabla _{\perp }$ in a
spherical system of coordinates

\begin{equation}
{\bf L}=\frac 1i\left( {\bf e}_\phi \frac \partial {\partial \theta }-\frac{%
{\bf e}_\theta }{\sin \theta }\frac \partial {\partial \phi }\right) ;{\bf %
\nabla }_{\perp }=\frac{{\bf e}_\theta }r\frac \partial {\partial \theta }+%
\frac{{\bf e}_\phi }{r\sin \theta }\frac \partial {\partial \phi },
\label{b19}
\end{equation}
we obtain

\begin{equation}
{\bf d\cdot B}=R\sum\limits_{l=1}^{\infty \vee
L}\sum\limits_{m=-l}^lc_l^m\left\{ \frac url(l+1)-\overline{\alpha }%
im+(1-u^2)\frac d{du}\frac 1r\frac d{dr}r\right\} f_l\widetilde{P}%
_l^{|m|}(u)e^{im\phi }.  \label{b20}
\end{equation}
Here ${\bf e}_\phi $ and ${\bf e}_\theta $ are unit vectors of the spherical
system of coordinates, and $u=\cos \theta $. To determine the coefficients
of the initial expansion we take advantage of the projection scheme from (%
\cite{Rudenko}). We multiply the equality (\ref{b20}) by the factor $%
\widetilde{P}_l^{|m|}\left( \cos \theta \right) e^{-im\phi }\cos \theta \sin
\theta d\phi d\theta $, and integrate it on a sphere $|{\bf r}|=R$ over
angular variables:

\begin{equation}
\begin{array}{l}
\int\limits_0^\pi \int\limits_0^{2\pi }B_d(\theta ,\varphi )\widetilde{P}%
_l^{|m|}\left( \cos \theta \right) e^{-im\phi }\cos \theta \sin \theta d\phi
d\theta = \\
\qquad 2\pi \sum\limits_{l=1}^{\infty \vee L}c_k^m\int\limits_{-1}^1\left\{
\left[ u^2k(k+1)-i\overline{\alpha }Rmu\right] \widetilde{P}_k^{|m|}\left(
u\right) \widetilde{P}_l^{|m|}\left( u\right) \right.  \\
\qquad \left. -u(u^2-1)g_k\frac{d\widetilde{P}_k^{|m|}}{du}\widetilde{P}%
_l^{|m|}\left( u\right) \right\} du
\end{array}
\label{b21}
\end{equation}
Here $g_k=\left[ \frac d{dr}rf_k(r)\right] _{|_{r=R}}$ It is also assumed
that $f_k(R)=1$, which can always be done. Using the orthogonality property
of Legendre polynomials and the known recurrence relations for expressions
of the form $uP_n^m$ and $(u^2-1)dP_n^m/du$ (\cite{Abramowitz and Stegun}),
and introducing the designation for the left-hand side of the equality (\ref
{b21}) $b_l^m$ (these quantities will be referred to as the weight
coefficients)we can bring the expression (\ref{b21}) into the form:

\begin{equation}
b_l^m=a_{-2}^{m,l}c_{l-2}^m+a_{-1}^{m,l}c_{l-1}^m+a_0^{m,l}c_l^m+a_{+1}^{m,l}c_{l+1}^m+a_{+2}^{m,l}c_{l+2}^m,
\label{b22}
\end{equation}
where

\begin{equation}
\begin{array}{l}
a_{-2}^{m,l}=\frac{w_{l-1}^m}{\sqrt{w_{l-2}^mw_l^m}}\frac{(l-m-1)(l+m)(l-2)}{%
(2l-3)(2l+1)}(l-1-g_{l-2}), \\
a_{-1}^{m,l}=-i\alpha Rm\sqrt{\frac{w_l^m}{w_{l-1}^m}}\frac{(l-m)}{(2l-1)},
\\
a_0^{m,l}=\frac{w_{l+1}^m}{w_l^m}\left( \frac{l-m-1}{2l+1}\right)
^2l(l+1-g_l)+\frac{w_{l-1}^m}{w_l^m}\left( \frac{l+m}{2l+1}\right)
^2(l+1)(l+g_l), \\
a_{+1}^{m,l}=-i\alpha Rm\sqrt{\frac{w_l^m}{w_{l+1}^m}}\frac{(l+m+1)}{(2l+3)},
\\
a_{+2}^{m,l}=\frac{w_{l+1}^m}{\sqrt{w_{l+2}^mw_l^m}}\frac{(l-m+1)(l+m+2)(l+3)%
}{(2l+5)(2l+1)}(l+2+g_{l+2}).
\end{array}
\label{b23}
\end{equation}
In view of the property of (\ref{b14}), when determining the coefficients $%
c_l^m$, it is sufficient to use equations (\ref{b22}) for positive $m$ only.
The equalities of (\ref{b22}) can be broken up into $m$ independent systems
of equations, each of which can be represented in the matrix form (for a
finite expansion to $L$)

\begin{equation}
A_{ij}^mC_j^m=B_i^m,  \label{b24}
\end{equation}
where

\begin{equation}
\begin{array}{l}
A_{ij}^m=\delta _{j+2}^ia_{_{-2}}^{m,2j+m-2}+\delta
_{j+1}^ia_{_{-1}}^{m,2j+m-2}+\delta _j^ia_0^{m,2j+m-2} \\
\qquad +\delta _{j-1}^ia_{_{+1}}^{m,2j+m-2}+\delta _{j-2}^ia_{+2}^{m,2j+m-2},
\\
C_j^m=c_{2j+m-2}^m,B_i^m=b_{2i+m-2}^m,1\leq i,j\leq (L-m)/2+1,\delta
_j^i=\left\{
\begin{array}{c}
1,i=j \\
0,i\neq j
\end{array}
\right.
\end{array}
\label{b25}
\end{equation}
Hence, by determining the coefficients c from equation (\ref{b24}), we solve
the problem formulated. For each value of $m$, the matrices $\widehat{A}^m$
in (\ref{b24}) have a pentadiagonal form. Since the selected values of $%
\alpha $ are always small and the behavior of the elements on even diagonals
is similar to the behavior of the respective elements of the matrix derived
in (\cite{Rudenko}), one might expect that calculations of this vector
equation will also present no problems.

\section{Conclusions}

The formulas obtained in this study for determining the expansion
coefficients of the magnetostatic solution can be used in further
investigation in calculations using real daily magnetograms. As shown in (%
\cite{Rudenko}), this technique can only work well with sufficiently high
resolution data, such as magnetograms from Kitt Peak Observatory.
Technically, the realization of the numerical code of the solution outlined
above is not too different in its complexity from the code of calculating an
analogous problem for the potential-field model. One might expect good
results from extrapolating solar plasma parameters throughout the visible
disk at the time of observation, to the heights of about the solar radius (%
\cite{Rudenko}).

\begin{acknowledgements}

I am grateful to Mr.~V.~G.~Mikhalkovsky for his assistance in preparing the
English version of this report.

This work was supported within the State Scientific Research program on
''Astronomy''as well as grant 00-02-16456 of the Russian Foundation
for Fundamental Research.

\end{acknowledgements}

\end{article}

\end{document}